\documentclass[aps,prb,twocolumn,superscriptaddress,showpacs]{revtex4}

\usepackage{graphicx}
\usepackage{dcolumn}
\usepackage{bm}

\begin{document}

\title{Heat transport of quasi-one-dimensional Ising-like
antiferromagnet BaCo$_2$V$_2$O$_8$ in the longitudinal and
transverse fields}

\author{Z. Y. Zhao}
\affiliation{Hefei National Laboratory for Physical Sciences at
Microscale, University of Science and Technology of China, Hefei,
Anhui 230026, People's Republic of China}

\author{X. G. Liu}
\affiliation{Hefei National Laboratory for Physical Sciences at
Microscale, University of Science and Technology of China, Hefei,
Anhui 230026, People's Republic of China}

\author{Z. Z. He}
\affiliation{Fujian Institute of Research on the Structure of Matter,
Chinese Academy of Sciences, Fuzhou, Fujian 350002, People's Republic of China}

\author{X. M. Wang}
\affiliation{Hefei National Laboratory for Physical Sciences at
Microscale, University of Science and Technology of China, Hefei,
Anhui 230026, People's Republic of China}

\author{C. Fan}
\affiliation{Hefei National Laboratory for Physical Sciences at
Microscale, University of Science and Technology of China, Hefei,
Anhui 230026, People's Republic of China}

\author{W. P. Ke}
\affiliation{Hefei National Laboratory for Physical Sciences at
Microscale, University of Science and Technology of China, Hefei,
Anhui 230026, People's Republic of China}

\author{Q. J. Li}
\affiliation{Hefei National Laboratory for Physical Sciences at
Microscale, University of Science and Technology of China, Hefei,
Anhui 230026, People's Republic of China}

\author{L. M. Chen}
\affiliation{Department of Physics, University of Science and
Technology of China, Hefei, Anhui 230026, People's Republic of
China}

\author{X. Zhao}
\affiliation{School of Physical Sciences, University of Science
and Technology of China, Hefei, Anhui 230026, People's Republic of
China}

\author{X. F. Sun}
\email{xfsun@ustc.edu.cn}

\affiliation{Hefei National Laboratory for Physical Sciences at
Microscale, University of Science and Technology of China, Hefei,
Anhui 230026, People's Republic of China}

\date{\today}

\begin{abstract}

The very-low-temperature thermal conductivity ($\kappa$) is
studied for BaCo$_2$V$_2$O$_8$, a quasi-one-dimensional Ising-like
antiferromagnet exhibiting an unusual magnetic-field-induced
order-to-disorder transition. The nearly isotropic transport in
the longitudinal field indicates that the magnetic excitations
scatter phonons rather than conduct heat. The field dependence of
$\kappa$ shows a sudden drop at $\sim$ 4 T, where the system
undergoes the transition from the N\'eel order to the
incommensurate state. Another dip at lower field of $\sim$ 3 T
indicates an unknown magnetic transition, which is likely due to
the spin-flop transition. Moreover, the $\kappa(H)$ in the
transverse field shows a very deep valley-like feature, which
moves slightly to higher field and becomes sharper upon lowering
the temperature. This indicates a magnetic transition induced by
the transverse field, which however is not predicted by the
present theories for this low-dimensional spin system.

\end{abstract}

\pacs{66.70.-f, 75.47.-m, 75.50.-y}

\maketitle

\section{Introduction}

Low-dimensional or frustrated quantum magnets are good candidates
for studying the quantum phase transition (QPT), an interesting
phenomenon in condensed matter driven by the strong quantum
fluctuations. QPT can be accessed by varying such parameters as
magnetic field, pressure, or doping concentration. Among them, the
magnetic-field-induced long-range order in the spin-gapped systems
has attracted much attention because it can be mapped into a
Bose-Einstein condensation of magnetic excitations.\cite{SG-1,
SG-2, SG-3, SG-4, SG-5, Sun_DTN} Contrary to the field-induced
ordering, a peculiar order-to-disorder transition in the presence
of magnetic field was recently discovered in BaCo$_2$V$_2$O$_8$
(BCVO),\cite{He-1, He-2, He-3} which is a quasi-one-dimensional
(1D) antiferromagnetic (AF) insulator.

BCVO crystallizes in the tetragonal $I41/acd$ space group, with
the edge-shared CoO$_6$ octahedra forming a four-step periodic
screw chain along the $c$ axis.\cite{Magnetization-1} The
low-temperature magnetic susceptibility indicated a large magnetic
anisotropy and the $c$ axis is the spin-easy
axis.\cite{He-1,He-2,He-3} The Hamiltonian of BCVO in the magnetic
field can be described by a 1D $S$ = 1/2 (Co$^{2+}$) $XXZ$
Heisenberg model:
\begin{displaymath}
\begin{array}{r@{~}l}
H =& J \displaystyle\sum_{i} \{S_{i,z}S_{i+1,z} +  \epsilon(S_{i,x}S_{i+1,x} + S_{i,y}S_{i+1,y})\} \\
\\
   &- \mu_{B}\displaystyle\sum_{i} S_{i}gH,
\end{array}
\end{displaymath}
where $J_{\parallel}/k_B = 65$ K, $g = 6.2$, $\epsilon = 0.46$ and
$J_{\perp}/k_B = 65$ K, $g = 2.95$, $\epsilon = 0.46$ for
longitudinal and transverse field, respectively, without
considering the next-nearest-neighboring
interaction.\cite{Magnetization-2} BCVO is therefore regarded as a
quasi-1D Ising-like spin-chain system. In zero field, owing to the
inevitable interchain interaction, a long-range AF order is formed
below $T_N = 5.4$ K and the ordered Co$^{2+}$ spins align
antiferromagnetically along the screw chain.\cite{Structure} When
the longitudinal field ($H \parallel c$) is applied, a gapless
Tomonaga-Luttinger liquid (TLL) state is formed as the N\'eel
state is destroyed.\cite{Magnetization-1, Magnetization-2} This
unusual field-induced order-to-disorder transition is driven by
the quantum fluctuations. Moreover, below 2 K, an incommensurate
(IC) order is established above a temperature-independent critical
field of $H_c$ = 3.9 T and is resulted from the development of the
long-range correlation in the TLL spin chain.\cite{Incommensurate,
New_phase} In the IC ordered state, the longitudinal components of
the adjacent Co$^{2+}$ moments direct antiparallel along the
chain.\cite{Incommensurate}

The phase diagram was believed to be much simpler for a transverse
field ($H \perp c$), in which a spin polarization transition
occurs at very high field.\cite{He-3, Magnetization-2} However,
the high-field magnetic susceptibility in both longitudinal and
transverse fields revealed some anomalies that could not fit to
these phase boundaries. Another first-order phase boundary, which
has never been detected in the thermodynamic quantities, was
recently probed by the ultrasound measurement in the IC region and
was supposed to originate from the orbital
ordering.\cite{Phase_diagram} Apparently, more careful
experimental investigations using various techniques are necessary
for revealing the precise phase diagrams of BCVO. Low-temperature
heat transport has recently been found to be very useful for
probing the ground states and phase transitions in the
low-dimensional spin systems.\cite{Sun_DTN, Kohama_DTN, Ando,
Sologubenko1, Sologubenko2, Ba3Mn2O8, MCCL} In this paper, we
study the very-low-temperature heat transport of BCVO in both
longitudinal and transverse fields. It is found that the magnetic
excitations play an important role in the heat transport
properties by scattering phonons, which causes drastic changes of
thermal conductivity across the magnetic phase transitions. In
particular, some anomalies of $\kappa(H)$ associated with unknown
phase boundaries are observed. Our results demonstrate that the
existing theories based on the 1D $S$ = 1/2 $XXZ$ Heisenberg
model\cite{theory-1, theory-2, theory-3, theory-4, theory-5,
theory-6} are still far from describing this spin system well.

\section{Experiments}

The high-quality BCVO single crystals were grown by a spontaneous
nucleation method.\cite{He-1} The crystals were accurately
oriented by using the x-ray back-scattering Laue photographs with
a precision of 1$^\circ$. Two samples with dimensions of $1.2
\times 0.59 \times 0.16$ mm$^3$ and $1.0 \times 0.67 \times 0.38$
mm$^3$ were used to measure the thermal conductivity along and
perpendicular to the spin-chain direction ($\kappa_c$ and
$\kappa_{ab}$), respectively. The thermal conductivity were
measured using a conventional steady-state technique and two
different processes: (i) using a ``one heater, two thermometers"
technique in a $^3$He refrigerator and a 14 T magnet at
temperature regime of 0.3 -- 8 K; (ii) using a Chromel-Constantan
thermocouple in a pulse-tube refrigerator for the zero-field data
above 4 K.\cite{Sun_DTN, HoMnO3, GdFeO3, Ba3Mn2O8, MCCL}

\section{Results and Discussion}

\begin{figure}
\includegraphics[clip,width=7.0cm]{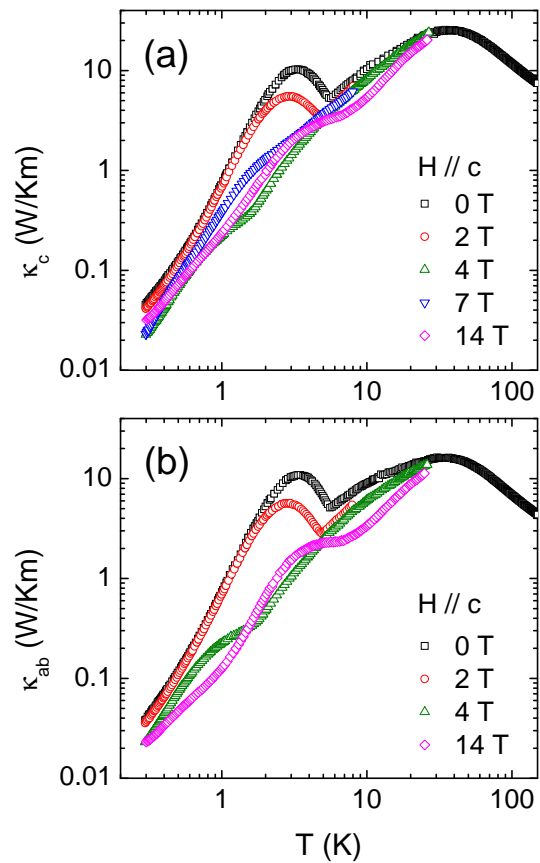}
\caption{(Color online) Temperature dependencies of thermal
conductivity of BaCo$_2$V$_2$O$_8$ single crystals in different
magnetic fields applied parallel to the $c$ axis with the heat
current parallel (a) and perpendicular (b) to the $c$ axis,
respectively.}
\end{figure}

Figure 1 shows the temperature dependencies of thermal
conductivity of BCVO single crystals in the longitudinal magnetic
field with the heat current parallel and perpendicular to the $c$
axis, respectively. It is clearly seen that no matter which
direction the heat flows, $\kappa$ exhibits similar behaviors
under the influence of longitudinal field. It is well known that
in the strongly anisotropic spin systems, such as the spin-chain
materials, the magnetic excitations are not able to propagate
perpendicular to the chain direction since the interchain spin
interaction is much weaker than the intrachain one. As a result,
$\kappa$ perpendicular to the spin-chain direction can only be the
contribution from the phonons, but $\kappa$ along the chain can be
the sum of phonons and magnetic excitations.\cite{Brenig, Hess,
Sologubenko3} However, the nearly isotropic behavior of $\kappa_c$
and $\kappa_{ab}$ and much weaker temperature dependence than
$T^3$ at very low temperatures indicate that the heat carriers are
mainly phonons for both directions of heat current. Therefore, the
magnetic excitations in BCVO could act as a kind of phonon
scatterers.

The zero-field $\kappa(T)$ curves show a double-peak structure
with a dip locating at $\sim$ 5.4 K, which is apparently related
to the N\'eel order of Co$^{2+}$ spins.\cite{Structure} Upon
increasing the magnetic field, the dip shifts to lower
temperature, in agreement with the specific-heat results that the
longitudinal field can suppress the AF order.\cite{He-2,He-3} The
double-peak structure is thus a result of strong phonon scattering
by magnetic fluctuations, which are most significant at the phase
transition. Moreover, the lower-$T$ peak weakens gradually due to
the strengthened quantum fluctuation with increasing the field
until 4 T, near the transition from N\'eel order to IC phase at
$H_c$,\cite{Incommensurate} at which field the lower-$T$ peak
evolutes into a shoulder. If the field is increased further, it
seems that the shoulder moves to higher temperature. It can also
be seen that $\kappa$ in the magnetic field is never larger than
the zero-field value even if a high field of 14 T is applied,
indicating that the field-induced quantum fluctuation is dominant
in the low temperature and high field region.

\begin{figure}
\includegraphics[clip,width=8.5cm]{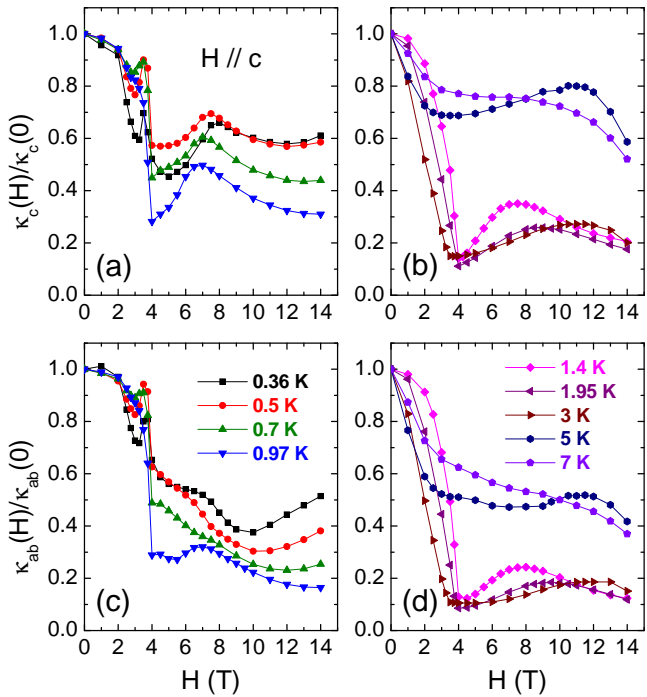}
\caption{(Color online) Magnetic-field dependencies of $\kappa_c$
(a,b) and $\kappa_{ab}$ (c,d) for field along the spin-chain
direction. }
\end{figure}

The magnetic field dependencies of thermal conductivity with
different heat-current directions in the longitudinal field are
shown in Fig. 2. Similar to the $\kappa(T)$ curves, the
$\kappa(H)$ isotherms also exhibit nearly isotropic transport
behavior, again indicating the phonon scattering by magnetic
excitations. In the presence of the longitudinal field, $\kappa$
is always smaller than the zero-field value, and the strongest
suppression of $\kappa$ is down to $\sim$ 10\%. The most
remarkable feature of $\kappa(H)$ is the steep drop at $H_c \sim$
4 T, and this drop becomes broader and deeper with increasing
temperature. Clearly, if the dip of $\kappa(H)$ corresponds to a
magnetic transition, it is coincided with the phase transition
from the N\'eel state to the IC state below 2 K or the one from
the N\'eel state to the disordered state above 2
K,\cite{Incommensurate} respectively. Additionally, there is
another dip around $\sim$ 3 T at subKelvin temperatures, and the
deepness of the dip is gradually reduced with increasing
temperature; for example, at $T = 0.97$ K the dip nearly
disappears, as shown in Figs. 2(a) and 2(c). This kind of anomaly
in a long-range AF state is most likely a consequence of the
spin-flop transition for the magnetic field applied along the
spin-easy axis.\cite{GdFeO3, HoMnO3} However, as mentioned before,
the spin system of BCVO was described by the $XXZ$ model with an
Ising-like anisotropy, which actually cannot allow the spin
flop.\cite{Magnetization-1} The present results therefore suggest
that the spin anisotropy is not strong enough to prevent the
occurrence of spin flop. As far as other experimental studies are
concerned, one may note that the magnetization, which can probe
the spin flop, has not yet been carried out down to subKelvin
temperatures.\cite{Magnetization-1, Magnetization-2}

Nevertheless, the magnetic field dependence of $\kappa$ can be
mainly ascribed to the scattering on phonons induced by the
quantum fluctuations, which are strengthened with increasing
field. The field-induced quantum fluctuations destroy the
long-range order but meanwhile establish the IC order. Across the
phase transition, the scattering of magnetic excitations is
pronounced and leads to a much more prominent feature in
$\kappa(H)$. With further increasing field, the quantum
fluctuations can even destroy the IC state and form a disordered
phase. A broad-peak-like behavior of $\kappa(H)$ in high fields
and at high temperatures may result from the complicated
interactions in the disordered phase.

\begin{figure}
\includegraphics[clip,width=7.0cm]{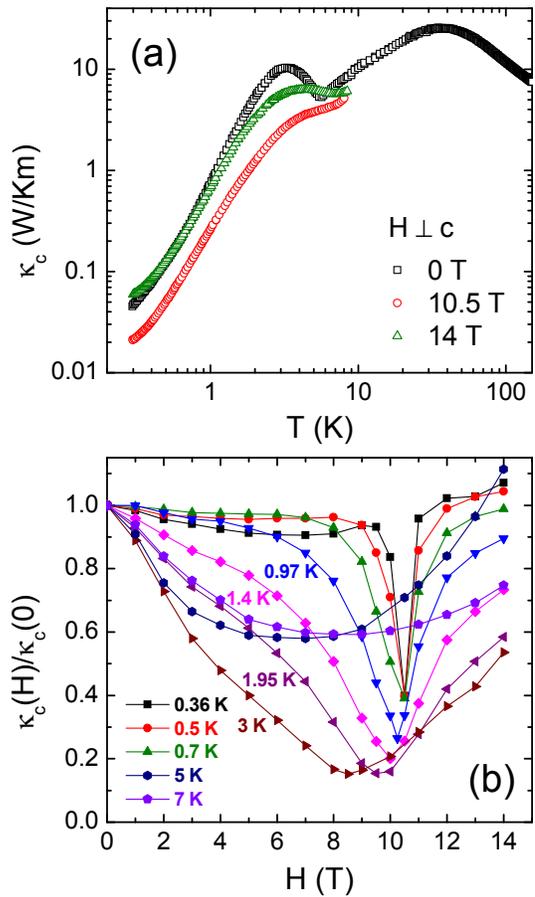}
\caption{(Color online) The $c$-axis (intrachain) thermal
conductivity of BaCo$_2$V$_2$O$_8$ single crystal versus
temperature (a) and magnetic field (b). The magnetic field is
applied perpendicular to the $c$ axis.}
\end{figure}

By now, we have a relatively comprehensive understanding of the
physical properties of BCVO in the longitudinal field based on
many different measurements. The heat transport data are in
general consistent with other results. Whereas the experimental
results about the effect of the transverse field are relatively
rare\cite{He-3, Magnetization-2} in spite of a lot of theoretical
works.\cite{theory-1, theory-2, theory-3, theory-4, theory-5,
theory-6} The temperature dependencies of intrachain thermal
conductivity in the transverse field are shown in Fig. 3(a).
Different from the case of the longitudinal field, in which the
conductivity in high field is much smaller than the zero-field
value, the conductivity in the transverse field is first
decreased, then followed by an enhancement with increasing field.
In 14 T field, the low-$T$ thermal conductivity is nearly equal to
the zero-field one, except for a weak reduction around the low-$T$
peak.

As shown in Fig. 3(b), the behavior of $\kappa_c(H)$ in the
transverse field is completely different from that in the
longitudinal field. It can be seen that, at very low temperatures,
the thermal conductivity is almost independent of the field except
for a minimum at $\sim$ 10 T. Upon increasing temperature, the
minimum becomes broader and deeper and shifts to lower field, and
finally evolutes into a shallow valley ranged over all the field
region. First of all, although the phonon resonant scattering by
paramagnetic moments could produce a similar dip-like $\kappa(H)$
behavior,\cite{PLCO, GdBaCoO} it is not likely the case in the
present work. One clear discrepancy between the phenomenon of Fig.
3 and paramagnetic scattering is that in the latter case the
minimum of $\kappa(H)$ shifts to higher field as the temperature
is increased.\cite{PLCO, GdBaCoO} Furthermore, the paramagnetic
scattering is closely related to the Zeeman effect and therefore
should be essentially isotropic on the field direction, which is
apparently different from what the BVCO data show. Therefore, it
is more likely that there is some kind of magnetic transition
responsible for the present results. The feature that the dip in
Fig. 3(b) becomes sharper at $T \rightarrow$ 0 manifests that this
transition is probably a QPT or spin-flop like. It is also notable
that the transition associated with this transport behavior is
different from the phase transition between the AF state and the
paramagnetic state under the transverse field, which can be probed
by the specific heat measurement.\cite{He-3} In particular, the
change of the AF transition temperature is not significant even if
the field is increased up to 9 T, whereas the change of the
minimum of $\kappa$ is much more sizable. If we plot the AF
transition points from the specific heat together with those from
the heat transport in the $H-T$ phase
diagram,\cite{demagnetization} as shown in Fig. 4, it can be seen
that they stand for two different phase boundaries. Apparently,
the transitions of the $\kappa(H)$ curves are likely associated
with some field-induced magnetic transition inside the AF state.

\begin{figure}
\includegraphics[clip,width=8.0cm]{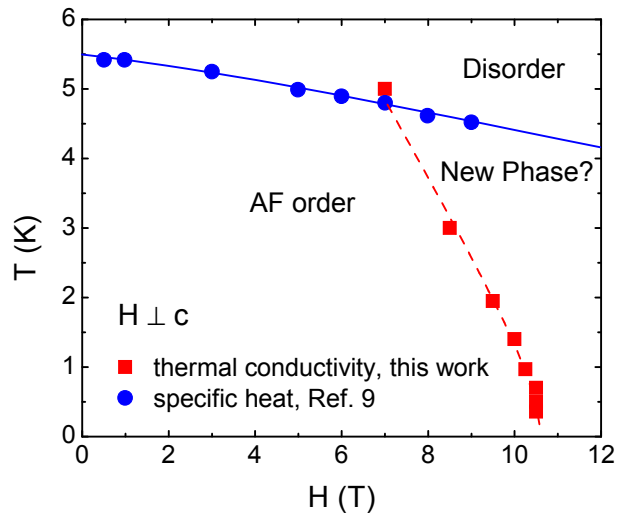}
\caption{(Color online) Magnetic phase diagram of
BaCo$_2$V$_2$O$_8$ for the transverse field $H \perp c$. The
transitions of the specific heat data (Ref. \onlinecite{He-3})
separate the long-range AF ordered state from the paramagnetic
disordered state. The transitions of the $\kappa_c(H)$ data
indicates that there is a new phase transition inside the AF
state. The lines are the guides for eyes.}
\end{figure}

Recently, Yamaguchi {\it et al.} found a similar magnetic
transition using the ultrasound measurement (for $H \perp c$), in
which some ``dip"-like anomalies appeared in the field dependence
of the sound velocity at nearly the same critical fields as those
in the $\kappa(H)$ curves.\cite{Yamaguchi} This possible phase
transition demonstrated by the heat transport and ultrasound
properties cannot, however, be explained by the existing theories
based on the Heisenberg $XXZ$ model. In this model, there is only
one possible transition in the transverse field, that is, from the
AF state to the paramagnetic state.\cite{theory-2, theory-3,
theory-4, theory-5} Note that the earlier high-field magnetization
data were actually also difficult to
understand.\cite{Magnetization-2} On the one hand, the high-field
magnetization at 1.3 K showed saturation at about 40 T, which is
in good agreement with the theoretical results. On the other hand,
it showed an anomaly at about 30 T, which is not expected from the
theories. (In passing, the even lower-$T$ magnetization would be
useful to reveal the anomaly of thermal conductivity at $\sim$ 10
T.) Therefore, these discrepancies between experiments and
theories, as well as the spin-flop-like transition in the
longitudinal field, demonstrate that the Heisenberg $XXZ$ model
may not be able to catch all the inherent properties of BCVO.
Probably, either the finite interchain spin exchange or the
deviation from the Ising anisotropy should be carefully considered
in a more legitimate model to describe the spin system of BCVO.

\section{Summary}

We have studied the heat transport of BaCo$_2$V$_2$O$_8$ single
crystal, which is a quasi-1D Ising-like $S$ = 1/2 AF compound. In
the longitudinal field, the nearly isotropic temperature
dependencies of $\kappa$ demonstrates the scattering of magnetic
excitations on phonons. There is a distinct and sharp decrease in
$\kappa(H)$ isotherms, which is related to the transition from the
N\'eel order to the IC state at $H_c \sim$ 4 T. Moreover, another
dip of $\kappa(H)$ at $\sim$ 3 T is supposed to result from the
spin-flop transition, which has never been observed before. On the
other hand, a possible magnetic transition is observed in the
transverse field, which is evidenced as a minimum in the
$\kappa(H)$ curve. This novel transition is consistent with a
recent ultrasound measurement, but its origin remained unclear and
there is clear need for more detailed investigations.

\begin{acknowledgements}

We thank H. Yamaguchi for sharing their unpublished data and Y.
Takano for helpful discussions. This work was supported by the
Chinese Academy of Sciences, the National Natural Science
Foundation of China, and the National Basic Research Program of
China (Grant Nos. 2009CB929502 and 2011CBA00111).

\end{acknowledgements}


\begin{thebibliography}{}
\bibitem{demagnetization}
The demagnetizing effect in this compound could be neglected. The
inner field of the crystal is expressed as $H_i = H_e/(1+n\chi)$,
where $H_e$ is the external magnetic field, $n$ is the
demagnetizing factor, and $\chi$ is the magnetic susceptibility.
Due to the antiferromagnetism at low temperatures, the correction
factor $n\chi$ is as small as 10$^{-2}$.
\end{thebibliography}
\end{document}